\documentclass[prd,a4paper,twocolumn]{revtex4}
\usepackage{graphicx}
\usepackage{subfigure,amssymb}

\newcommand{\nc}{\newcommand}

\nc{\be}[1]{\begin{equation}\mbox{$\label{#1}$}}
\nc{\bea}[1]{\begin{eqnarray} \mbox{$\label{#1}$}}
\nc{\Section}[2]{\section{#2}\label{#1}}
\nc{\Bibitem}[1]{\bibitem{#1}}
\nc{\Label}[1]{\label{#1}}

\nc{\eea}{\end{eqnarray}}
\nc{\ee}{\end{equation}}

\nc{\bdm}{\begin{displaymath}}
\nc{\edm}{\end{displaymath}}
\nc{\dpsty}{\displaystyle}
\nc{\bc}{\begin{center}}
\nc{\ec}{\end{center}}
\nc{\ba}{\begin{array}}
\nc{\ea}{\end{array}}
\nc{\bab}{\begin{abstract}}
\nc{\eab}{\end{abstract}}
\nc{\btab}{\begin{tabular}}
\nc{\etab}{\end{tabular}}
\nc{\bit}{\begin{itemize}}
\nc{\eit}{\end{itemize}}
\nc{\ben}{\begin{enumerate}}
\nc{\een}{\end{enumerate}}
\nc{\bfig}{\begin{figure}}
\nc{\efig}{\end{figure}}

\nc{\arreq}{&\!=\!&}
\nc{\arrmi}{&\!-\!&}
\nc{\arrpl}{&\!+\!&}
\nc{\arrap}{&\!\!\!\approx\!\!\!&}
\nc{\non}{\nonumber}
\nc{\align}{\!\!\!\!\!\!\!\!&&}

\def\lsim{\; \raise0.3ex\hbox{$<$\kern-0.75em
      \raise-1.1ex\hbox{$\sim$}}\; }
\def\gsim{\; \raise0.3ex\hbox{$>$\kern-0.75em
      \raise-1.1ex\hbox{$\sim$}}\; }

\nc{\DOT}{\hspace{-0.08in}{\bf .}\hspace{0.1in}}
\nc{\Laada}{\hbox {$\sqcap$ \kern -1em $\sqcup$}}
\nc\loota{{\scriptstyle\sqcap\kern-0.55em\hbox{$\scriptstyle\sqcup$}}}
\nc\Loota{{\sqcap\kern-0.65em\hbox{$\sqcup$}}}
\nc\laada{\Loota}
\nc{\qed}{\hskip 3em \hbox{\BOX} \vskip 2ex}

\nc{\real}{{\rm I \! R}}
\nc{\Z}{{\sf Z \!\!\! Z}}
\nc{\complex}{{\rm C\!\!\! {\sf I}\,\,}}
\def\bigid{\leavevmode\hbox{\small1\kern-3.8pt\normalsize1}}
\def\id{\leavevmode\hbox{\small1\kern-3.3pt\normalsize1}}
\nc{\slask}{\!\!\!/}
\nc{\bis}{{\prime\prime}}
\nc{\pa}{\partial}
\nc{\na}{\nabla}
\nc{\ra}{\rangle}
\nc{\la}{\langle}
\nc{\goto}{\rightarrow}
\nc{\swap}{\leftrightarrow}

\nc{\EE}[1]{ \mbox{$\cdot10^{#1}$} }
\nc{\abs}[1]{\left|#1\right|}
\nc{\at}[2]{\left.#1\right|_{#2}}
\nc{\norm}[1]{\|#1\|}
\nc{\abscut}[2]{\Abs{#1}_{\scriptscriptstyle#2}}
\nc{\vek}[1]{{\rm\bf #1}}
\nc{\integral}[2]{\int\limits_{#1}^{#2}}
\nc{\inv}[1]{\frac{1}{#1}}
\nc{\dd}[2]{{{\partial #1}\over{\partial #2}}}
\nc{\ddd}[2]{{{{\partial}^2 #1}\over{\partial {#2}^2}}}
\nc{\dddd}[3]{{{{\partial}^2 #1}\over
    {\partial #2 \partial #3}}}
\nc{\dder}[2]{{{d #1}\over{d #2}}}
\nc{\ddder}[2]{{{d^2 #1}\over{d {#2}^2}}}
\nc{\dddder}[3]{{d^2 #1}\over
    {d #2 d #3}}
\nc{\dx}[1]{d\,^{#1}x}
\nc{\dy}[1]{d\,^{#1}y}
\nc{\dz}[1]{d\,^{#1}z}
\nc{\dl}[1]{\frac{d\,^{#1}l}{(2\pi)^{#1}}}
\nc{\dk}[1]{\frac{d\,^{#1}k}{(2\pi)^{#1}}}
\nc{\dq}[1]{\frac{d\,^{#1}q}{(2\pi)^{#1}}}

\nc{\bfT}{{\bf T }}

\nc{\cA}{{\cal A}}
\nc{\cB}{{\cal B}}
\nc{\cD}{{\cal D}}
\nc{\cE}{{\cal E}}
\nc{\cG}{{\cal G}}
\nc{\cH}{{\cal H}}
\nc{\cL}{{\cal L}}
\nc{\cO}{{\cal O}}
\nc{\cT}{{\cal T}}
\nc{\cN}{{\cal N}}
\nc{\cR}{{\cal R}}
\nc{\rvac}[1]{|{\cal O}#1\rangle}
\nc{\lvac}[1]{\langle{\cal O}#1|}
\nc{\rvacb}[1]{|{\cal O}_\beta #1\rangle}
\nc{\lvacb}[1]{\langle{\cal O}_\beta #1 |}
\nc{\bb}{\bar{\beta}}
\nc{\bt}{\tilde{\beta}}
\nc{\ctH}{\tilde{\cal H}}
\nc{\chH}{\hat{\cal H}}
\nc{\1}{\aa}
\nc{\2}{\"{a}}
\nc{\3}{\"{o}}
\nc{\4}{\AA}
\nc{\5}{\"{A}}
\nc{\6}{\"{O}}
\nc{\al}{\alpha}
\nc{\g}{\gamma}
\nc{\Del}{\Delta}
\nc{\e}{\textrm{e}}
\nc{\eps}{\epsilon}
\nc{\lam}{\lambda}
\nc{\Om}{\Omega}
\nc{\ve}{\varepsilon}
\nc{\mn}{{\mu\nu}}
\nc{\vp}{\varphi}

\nc{\rf}[1]{(\ref{#1})}
\nc{\nn}{\nonumber \\*}
\nc{\bfB}{\bf{B}}
\nc{\bfv}{\bf{v}}
\nc{\bfx}{\bf{x}}
\nc{\bfy}{\bf{y}}
\nc{\vx}{\vec{x}}
\nc{\vy}{\vec{y}}
\nc{\oB}{\overline{B}}
\nc{\oI}{\overline{I}}
\nc{\oR}{\overline{R}}
\nc{\rar}{\rightarrow}
\nc{\ti}{\times}
\nc{\slsh}{\hskip-5pt/}
\nc{\sm}{Standard~Model~}
\nc{\MP}{M_{\rm Pl}}
\nc{\mpl}{M_{\rm Pl}}
\nc{\tp}{t_{\rm Pl}}

\nc{\pmin}{p_{\rm min}}
\nc{\pmax}{p_{\rm max}}
\nc{\fo}{f_0}
\nc{\foi}{f_{0,i}\,}
\nc{\fop}{f_0^P}
\nc{\fou}{f_0^U}

\nc{\eff}{{\rm eff}}
\nc{\MT}{M_{\rm T}}
\nc{\ML}{M_{\rm L}}
\nc{\kk}{\vek{k}}
\nc{\pp}{{\rm p}}
\nc{\pt}{\partial_t}
\nc{\half}{{1\over 2}}
\nc{\w}{\omega}
\nc{\uhat}{\hat{U}_\w}

\nc{\etal}{\mbox{\it et al.}}
\nc{\ie}{{\it i.e. }}
\nc{\eg}{{\it e.g. }}
\nc{\trh}{T_{\rm RH}}
\nc{\ad}{{a'\over a}}
\nc{\bd}{{b'\over b}}
\nc{\Rd}{{R'\over R}}
\nc{\diag}{{\textrm{diag}}}
\nc{\mato}[1]{\tilde{#1}}
\nc{\sech}{\textrm{sech}}
\nc{\I}{\textrm{I}}
\nc{\II}{\textrm{II}}
\nc{\III}{\textrm{III}}
\nc{\vev}[1]{\langle #1 \rangle}
\nc{\hyp}{\,\; F_{1{\hskip -16pt}2}{\hskip 11pt}}
\nc{\brhom}{\overline{\rho}_M}
\nc{\brho}{\overline{\rho}}
\nc{\rhob}{\overline{\rho}}
\nc{\Pb}{\overline{P}}
\nc{\bH}{\overline{H}}
\nc{\ep}{{1+4\eps}}

\nc{\lcdm}{$\Lambda$CDM}

\def\smiley{\hbox{\large$\bigcirc$\hspace{-.80em}%
\raise.2ex\hbox{$\cdot\cdot$}\kern-.61em    %--- .56
\lower.2ex\hbox{\scriptsize$\smile$}}\ }

\def\frowney{\hbox{\large$\bigcirc$\hspace{-.80em}%
\raise.2ex\hbox{$\cdot\cdot$}\kern-.635em
\lower.2ex\hbox{\scriptsize$\frown$}}\ }

\begin{document}

\title{Static spherically symmetric perfect fluid solutions in $f(R)$ theories of gravity}

\author{T. Multam\"aki}
\email{tuomul@utu.fi}
\author{I. Vilja}
\email{vilja@utu.fi}
\affiliation{Department of Physics, University of Turku, FIN-20014 Turku, FINLAND}

\date{}

\begin{abstract}
Static spherically symmetric perfect fluid solutions are studied in metric $f(R)$ 
theories of gravity. We show that pressure and density do not uniquely
determine $f(R)$ {\it ie.} given a matter distribution and an equation state, 
one cannot determine the functional form of $f(R)$. However, we also show that matching 
the outside Schwarzschild-de Sitter-metric to the metric inside the mass distribution 
leads to additional constraints that severely limit the allowed fluid configurations. 
\end{abstract}

\maketitle

\section{Introduction}

Modern day cosmological observations such as supernovae type Ia \cite{snia},
cosmic microwave background \cite{cmb} and large scale structure 
\cite{lss} provide strong evidence against a critical density matter dominated
universe. Instead, the current cosmological concordance model is a critical density universe 
dominated by cold dark matter and dark energy in the form of some kind 
effective cosmological constant.
The traditional cosmological constant 
%introduced by Einstein himself
is maybe the leading dark energy candidate (for a review see {\it e.g.} 
\cite{peebles}), but a large number of other alternatives have been studied in the vast
literature on dark energy.

Modifying general relativity (GR) to explain the present day acceleration is an often considered
avenue of research. In particular, a class of models that has been extensively studied
in the recent years are the $f(R)$ gravity models that replace the Einstein-Hilbert (EH)-action
with an arbitrary function of the curvature scalar 
(see \eg \cite{turner,turner2,allemandi,meng,nojiri3,nojiri2,cappo1,woodard,odintsov} 
 and references therein).
Modifying the gravitational action is faced with many challenges, however, 
and obstacles such as instabilities \cite{dolgov,soussa,faraoni} 
as well as constraints arising from known properties of gravity
in our solar system (see {\it e.g.} \cite{chiba,confprobs,Clifton} and references therein)
need to be overcome. Also the large scale perturbations present a challenge
for $f(R)$ gravity theories \cite{Bean:2006up,Song:2006ej}.
One of the most direct and strictest constraints on any modification of gravity comes
from observations of our solar system. This is often done conformally transforming the 
theory to a scalar-tensor theory and then considering the Parameterized Post-Newtonian (PPN) limit 
\cite{damour,magnano} (see also \cite{olmo, ppnok} for a discussion).
The question of Solar System constraints on $f(R)$-theories has recently been
extensively discussed by a number of authors \cite{Erickcek:2006vf,Chiba2,Jin:2006if,
Faulkner:2006ub}.

The essence of the discussion is the validity of the Schwarzschild-de Sitter (SdS)
-solution in the Solar System. The SdS-metric is a vacuum solution to a large class
of $f(R)$-theories of gravity. However, due to the higher-derivative nature
of the metric $f(R)$-theories, it is not unique
and other solutions can also be constructed (see {\it eg.} \cite{cognola,Multamaki2}).
This fact is also present in the cosmological setting, rendering any cosmological solution 
non-unique and hence the form of $f(R)$ cannot be uniquely determined from the 
expansion history of the universe alone \cite{Multamaki}.

In the recent literature this question has now been addressed without resorting to scalar-tensor
theory \cite{Erickcek:2006vf,Chiba2,Jin:2006if}.
The result is compatible with the scalar-tensor theory calculations:
the Solar System constraints are valid in a particular limit that 
corresponds to the limit of light effective scalar in the equivalent scalar-tensor
theory.  In terms of the $f(R)$-theory, this is
equivalent to requiring that one can approximate
the trace of the field equations by Laplace's equation \cite{Chiba2}.
As a result, the often considered $1/R$ theory is not
consistent with the Solar System constraints in this limit, if the $1/R$ term is to 
drive late time cosmological acceleration.

In this work, we approach the question differently by asking what kind of a
mass distribution is required so that the SdS-metric is the solution outside
a spherically symmetric body. In other words, given the SdS-metric, what
is the matter distribution that has the correct boundary conditions.
Our approach is general, we do not make any assumptions about the 
$f(R)$-theory. 

Here we show that like in the cosmological setting, the mass distribution alone
cannot in general determine the gravitational theory, or the functional form of $f(R)$.
Imposing the SdS-metric as a boundary condition does limit the allowed solutions
however. We also give a prescription how one can in principle solve the mass distribution
that has the SdS-metric as the outside solution, given a gravitational theory and 
the equation of state of matter.

\section{$f(R)$ gravity formalism}

The action for $f(R)$ gravity is ($8\pi G=1$) 
\be{action}
S = \int{d^4x\,\sqrt{-g}\Big(f(R)+{\cal{L}}_{m}\Big)}.
\ee
The field equations resulting in the so-called metric approach are
reached by variating with respect to $g_{\mu\nu}$: 
\be{eequs}
F(R) R_{\mu\nu}-\frac 12 f(R) g_{\mu\nu}-\nabla_\mu\nabla_\nu F(R)+g_{\mu\nu}\Box F(R)=T^m_{\mu\nu},
\ee
where $T_{\mu\nu}^m$ is the standard minimally coupled stress-energy tensor
and $F(R)\equiv df/dR$.
Alternatively, this can be written in a form similar to the field equations of 
General Relativity (GR):
\be{eequs2}
G_{\mu\nu}\equiv R_{\mu\nu}-\frac 12 R g_{\mu\nu}=T^c_{\mu\nu}+\tilde T^m_{\mu\nu},
\ee
where the stress-energy tensor of the gravitational fluid is
\bea{tmunu}
T^c_{\mu\nu} & = & \frac{1}{F(R)}\Big\{\frac 12 g_{\mu\nu}\Big(f(R)-RF(R)
\Big)+\nonumber \\
& + & F(R)^{;\alpha\beta}\Big(g_{\alpha\mu}g_{\beta\nu}-g_{\mu\nu}
g_{\alpha\beta}\Big)\Big\}
\eea
and we have defined
\be{matter}
\tilde T^m_{\mu\nu}\equiv T^m_{\mu\nu}/F(R).
\ee

Contracting Eq. (\ref{tmunu}) and assuming that we can describe the
stress-energy tensor with a perfect fluid, we get
\be{contra}
F(R)R-2 f(R)+3\Box F(R)=\rho-3p.
\ee

\section{Spherically symmetric perfect fluid solutions}

We are interested in static, spherically symmetric
perfect fluid solutions, or as commonly referred to in the literature, SSSPF-solutions.
Study of such solutions can be traced back to Schwarzschild \cite{karl} and the 
literature is extensive (see {\it eg.} \cite{ssspf} for reviews). The surface of the fluid
sphere is defined by the surface of zero pressure where the interior solution
is matched to the outside metric.

In spherically symmetric coordinates the metric can generally be written as
\be{sphersym}
g_{\mu\nu}=\left(\begin{array}{cccc}
A(r) & 0 & 0 & 0\\
0 & B(r) & 0 & 0\\
0 & 0 & -r^2 & 0\\
0 & 0 & 0 & -r^2 \sin^2(\theta)
\end{array}
\right).
\ee
The corresponding continuity equation is
\be{cont}
\frac{p'(r)}{\rho(r)+p(r)}=-\frac 12 \frac{A'(r)}{A(r)},
\ee 
where prime indicates a derivation with respect to $r$, $'\equiv d/dr$.

\subsection{Uniqueness}

The higher derivative nature of metric $f(R)$ theories of gravity can lead to
non-uniqueness of solutions of the field equations. For example, the 
cosmological solution $a(t)$ for a given $f(R)$ theory is not unique but
other $f(R)$ theories with the same expansion history exists \cite{Multamaki}.
This also true for vacuum solutions: even though the SdS metric is a solution
of the field equations in vacuum, others also exist \cite{Multamaki2}.

Here we consider the question of uniqueness in the presence of matter.
This is most conveniently done by using a form of the field equations where
$f(R)$ is eliminated from the equations by using the contracted equation
Eq. (\ref{contra}):
\be{field2}
F R_{\mu\nu}-\frac 14 (FR-\Box F)g_{\mu\nu}-\nabla_\mu\nabla_\nu F=
T^m_{\mu\nu}-\frac 14(\rho-3p)g_{\mu\nu}.
\ee
Given a matter distribution, $\rho(r),\ p(r)$, one can 
solve for $A$ from the continuity equation and substitute into Eq. (\ref{field2})
giving a set of differential equations for $F(r)$ and $B(r)$.

This set can be solved algebraically for $B(r)$ and $B'(r)$ so that
both $B(r)$ and $B'(r)$ can be expressed in terms of $F(r),\ \rho(r),\ p(r)$
and their derivatives (see Appendix for details).
Differentiating the expression for $B(r)$ and equating it with the other expression
for $B'(r)$ we obtain
%Obviously $B(r)$ and $B'(r)$ are not independent and requiring that
%$d(B(r))/dr=B'(r)$ gives 
a single equation relating various derivatives of $F, \rho$ and $p$.
This equation is a non-linear third order differential equation for $F(r)$
and due to its length is not explicitly shown here. Here we adopt 
a short-hand notation for the equation and write
\be{mtov}
mTOV(F,\rho,\rho',\rho'',p,p',p'',p''')=S_f,
\ee
where $S_f=S_f(F',F'',F''',\rho,\rho',\rho'',p,p',p'',p''')$ represents a source term.
Explicit calculations show that in the GR limit, $F=1$, 
the source term vanishes, $S_f=0$, and Eq. (\ref{mtov})
is satisfied, whenever $\rho$ and $p$ satisfy the usual Tolman-Oppenheimer-Volkov (TOV)
-equation. Hence we can view Eq.\ (\ref{mtov}) as a modified TOV equation
of metric $f(R)$ theories of modified gravity. 

Given $\rho(r)$ and $p(r)$, one can solve Eq. (\ref{mtov}) for $F(r)$
and hence one has a solution for $R(r)$ using the expressions for $B(r)$
in terms of $F$. From $F(r)$ and $R(r)$ one can determine, at least in principle,
$F(R)$ and finally $f(R)$. The higher derivative of nature of $f(R)$ theories
is apparent in that Eq. (\ref{mtov}) can have a number of solutions. For example,
even when matter follows the ordinary TOV-equation, one can find non-trivial solutions
for $F(r)$ \ie for given standard SSSPF-solution of general relativity $F(r)=1$ is not
the only possible solution. 
%As a trivial example, considerwith which Eq. (\ref{mtov}) has
%a solution $F(r)=r$, in addition to the $F(r)=1$ solution. The non-trivial solution
%corresponds to $R(r)=...$, and hence the gravitational theory that has the Tolman ??
%solution as an exact solution is $f(R)=...$. This example demonstrates concretely
%how the gravitational theory is not uniquely determined by the matter distribution.

\subsection{Boundary conditions}

The outside solution sets the boundary conditions for the metric
components at the surface of the star, $r=r_0$. The field equations are 
fourth order in $A$ and third order in $B$ so that 
$A_0,\ A'_0\, A''_0,\ A'''_0,\ B_0,\ B_0',\ B_0''$, where $A_0\equiv A(r_0)$ etc.,
are fixed. This is in contrast to general relativity, where only $A_0,\ A_0',\ B_0$
are fixed by the outside solution. 

In this paper we are interested in solutions which have the 
Schwarzschild-de Sitter -space time (SdS) as the outside solution.
The SdS-metric is
\be{schwlam}
g_{\mu\nu}=\left(\begin{array}{cccc}
s(r)& 0 & 0 & 0\\
0 & -1/s(r) & 0 & 0\\
0 & 0 & -r^2 & 0\\
0 & 0 & 0 & -r^2 \sin^2(\theta)
\end{array}
\right),
\ee
where $s(r)\equiv 1-2M/r+r^2/12 R_0$ and $R_0$ is the corresponding 
scalar curvature (in the standard SdS representation
$R_0=-4\Lambda$). In the $f(R)$-theory framework $R_0$ is set by the equation 
$R_0 F(R_0)=2 f(R_0)$.

This choice of metric fixes 
the curvature scalar and its first derivative at the stellar surface at $r_0$: $R(r_0)=R_0$, 
$R'(r_0)=0$.
The surface of the star is set by the conditions$p(r_0)=0$ while the energy density
is left as a free parameter, $\rho(r_0)=\rho_0$.

Using these boundary conditions in the modified Einstein's equations, (\ref{eequs}),
along with the requirement $R_0 F(R_0)=2 f(R_0)$, gives two independent equations:
\bea{help}
s(r)\Big(\rho_0+s(r)\frac{d^2f}{dR^2}\Big |_{R_0}R''\Big)&=&0,\nonumber\\ 
s(r)\frac{d^2f}{dR^2}\Big |_{R_0}R''&=&0. 
\eea
Since we are interested in generalized gravity,
$df^2/dR^2|_{R_0}\neq 0$ and hence $R''(r_0)=0,\ \rho_0=0$. Note that this is already
a result different from general relativity where $\rho$ can be discontinuous 
at the surface.

Derivating the modified Einstein's equations once with respect to $r$ and applying the boundary
conditions along with the new constraints $R''(r_0)=0=\rho_0$, 
gives $R'''(r_0)=0,\ \rho'(r_0)=0$. From the continuity equation it then
straightforwardly follows that also $p''(r_0)=0$.
This process can then be continued by derivating the modified Einstein's equations 
once more and substituting the boundary values found at previous steps, resulting
in three independent equations for the highest order derivatives of $R,\ A$ and $B$.
These can then be solved and one can proceed to the next order derivatives. 
Obviously, derivatives of $R$ and $A$ and $B$ are not independent but no new information
is obtained from their mutual relation.

Higher derivatives of $\rho$ and $p$ are not uniquely defined and one is free to 
choose $\rho''$ and higher derivatives on the surface in a way that
reflects the desired equation of state.

In summary, requiring that the outside metric is the SdS-metric, sets
$\rho(r_0)=\rho'(r_0)=p(r_0)=p'(r_0)=p''(r_0)=0$ on the surface for a general $f(R)$ 
theory for which $d^2f/dR^2|_{R_0}\neq 0$. We can already see here how
the higher-derivative nature of $f(R)$ theories lead to a more
constrained system than general relativity.

\subsection{Polytropic stars}

As a simple example of how the boundary conditions already limit the range of allowed
solutions, we briefly consider polytropic stars.
White dwarfs and neutron starts are often approximated by a polytropic
equation of state, $p=\kappa \rho^\gamma$, where $\kappa$ and $\gamma=1+1/n$
are constants and $n$ is often referred to as the polytropic index
(see {\it eg.} \cite{weinberg2}).

The continuity equation, Eq. (\ref{cont}),
can straightforwardly be solved for such an equation of state:
\be{polysol}
\rho=\kappa^{1/(1-\gamma)}\Big((\frac{B}{B_0})^{(1-\gamma)/(2\gamma)}-1\Big)^{1/(\gamma-1)},
\ee
where we have used the fact that $\rho$ vanishes at the surface. Clearly this condition
sets that $\gamma>1$. Similarly, since $\rho'$ is also vanishing
at the boundary, we must require that $\gamma<2$. This is again quite different from 
GR, where $\gamma$ is unbounded from above (again see {\it eg.} \cite{weinberg2}).

\subsection{SSSPF with a SdS boundary metric}

In light of the discussion presented above, we can now consider perfect fluid
matter in $f(R)$ gravity theories with the SdS metric as a boundary condition.
Again, it is advantageous to consider a form of the field equations where
explicit $f(R)$ dependence is eliminated in favor of $F(r)$, Eq. (\ref{field2}).
Since $R'(r_0)=R''(r_0)=R'''(r_0)=0$, it clear that  $F'(r),\ F''(r),\ F'''(r)$ vanish
on the boundary {\it ie.} $F'(r_0)=F''(r_0)=F'''(r_0)=0$. The boundary value of $F$
can be solved from the requirement $R_0 F(r_0)=2 f(R_0)$ once $f(R)$ is given, 
{\it eg.}
if $f(R)=R-\mu^4/R$, $R_0^2=3\mu^4$ and $F_0=F(R_0)=4/3$.
Note that with a more general choice of $f(R)$, one can in principle easily mimic
general relativity {\it ie.} $F_0=1$ everywhere. However, requiring
$R_0 F_0=2 f(R_0)$ along with Eq. (\ref{contra}), implies that $\rho-3p=0$
everywhere inside the star making such constructions physically uninteresting.

Given a fluid sphere, $\rho(r),\ p(r)$ with appropriate boundary conditions,
one can in principle solve Eq. (\ref{mtov}) with the aforementioned boundary values for $F(r)$.
Using the solution $f(R)$ can then be reconstructed. The constant solution, $F(r)=F_0=1$,
is a solution exactly when $\rho(r)$ and $p(r)$ satisfy the TOV-equation
\ie matter is distributed like in general relativity. {\it Vice versa}, if matter does not obey
TOV-equation, the $S_f$ term acts as source for the differential equation of $F(r)$ and thus
the boundary conditions are not strong enough to force $F$ be constant, 
but the solution is more general. 
An important restriction is, however, the boundary conditions for $\rho$ and $p$
making the set of allowed general relativistic SSSPF-solutions more restricted in $f(R)$ gravity, 
{\it eg.} the standard Shwarzschild fluid sphere with constant density is not allowed
since $\rho_0\neq 0$.

\section{Conclusions}

In the present letter we have discussed spherically symmetric solutions with non-trivial
matter distributions applicable to stellar systems in particular. Although we have restricted our
analysis to a simplified system described by a perfect fluid, similar conclusions are expected
to apply in more realistic cases. 

We find that, like in the cosmological case, the distribution of matter does not determine
the gravitational theory uniquely but due to the higher-derivative nature of the field equations,
different gravitational theories can support the same solution. Given the matter distribution,
$\rho(r),\ p(r)$, one can, at least in principle, construct a gravitational theory that has the
desired solution by solving the modified TOV-equation, Eq. (\ref{mtov}).

By considering configurations that are matched to a Schwarzschild-de Sitter metric, we find
that such configurations are more tightly constrained than those of general relativity. Again,
this is due to the higher derivative nature of the metric $f(R)$ theories of gravity that
requires matching of higher order derivatives at the boundary of the fluid sphere than in 
general relativity.

As a result, we find that stellar configurations, approximated by a perfect fluid sphere,
can be accommodated to an external SdS-solution, whenever $\rho,\ p$ and $f(R)$ are related by a 
the modified TOV equation. The conventional TOV-equations correspond exactly to the choice 
$f(R)\equiv R$, and departures from the standard TOV-equations necessarily lead to a more general
 gravitational action. This phenomenon maybe noteworthy whenever modifications to the Einstein-Hilbert 
action, $f(R)=R$, are small. Small changes to the EH-action are likely to lead only small modifications 
to stellar models, {\it ie.} density and pressure of the matter may deviate only slightly from the 
ordinary TOV-relation. On this basis it hence seems possible that realistic stellar models may be 
constructed also in $f(R)$ gravity models without violating constraints from the solar system. 
The exact nature and whether such solutions correspond to $f(R)$ theories that can act as dark energy 
requires more extensive analysis.
%To make more definite conclusions, more extensive analysis is most definitely needed. 

\acknowledgments
TM is supported by the Academy of Finland.

\begin{appendix}
\section*{Appendix}
\begin{widetext}
The set of modified field equations, Eq. (\ref{field2}), for a known matter
distribution in terms of $F(r)$ reads:
\bea{fbset}
0 & = & \frac{F}{2\,r^2} - \frac{3\,p}{4} - \frac{3\,\rho}{4} + 
  \frac{F\,{p}^2}{2\,r^2\,B\,{\left( p + \rho \right) }^2} + 
  \frac{F\,p\,\rho}
   {r^2\,B\,{\left( p + \rho \right) }^2} + 
  \frac{F\,{\rho}^2}
   {2\,r^2\,B\,{\left( p + \rho \right) }^2} - 
  \frac{F\,B'}{2\,r\,{B}^2} + \frac{F'}{2\,r\,B} - 
  \frac{B'\,F'}{8\,{B}^2} + 
  \frac{F\,p\,p'}{r\,B\,{\left( p + \rho \right) }^2}\nonumber\\
& & + \frac{F\,\rho\,p'}{r\,B\,{\left( p + \rho \right) }^2} - 
  \frac{F\,B'\,p'}{4\,{B}^2\,\left( p + \rho \right) } + 
  \frac{3\,F'\,p'}{4\,B\,\left( p + \rho \right) } - 
  \frac{F\,{p'}^2}{B\,{\left( p + \rho \right) }^2} - 
  \frac{F\,p'\,\rho'}
   {2\,B\,{\left( p + \rho \right) }^2} + \frac{F''}{4\,B} + 
  \frac{F\,p\,p''}{2\,B\,{\left( p + \rho \right) }^2} + 
  \frac{F\,\rho\,p''}{2\,B\,{\left( p + \rho \right) }^2}\nonumber\\
0 & = & 
\frac{F}{2\,r^2} + \frac{F}{2\,r^2\,B} + \frac{p}{4} + 
  \frac{\rho}{4} + \frac{F\,B'}{2\,r\,{B}^2} + 
  \frac{F'}{2\,r\,B} + \frac{3\,B'\,F'}{8\,{B}^2} - 
  \frac{F\,p\,p'}{r\,B\,{\left( p + \rho \right) }^2} - 
  \frac{F\,\rho\,p'}{r\,B\,{\left( p + \rho \right) }^2} - 
  \frac{F\,B'\,p'}{4\,{B}^2\,\left( p + \rho \right) }
\nonumber\\
& &  - 
  \frac{F'\,p'}{4\,B\,\left( p + \rho \right) } - 
  \frac{F\,{p'}^2}{B\,{\left( p + \rho \right) }^2}
 - \frac{F\,p'\,\rho'}
   {2\,B\,{\left( p + \rho \right) }^2} - 
  \frac{3\,F''}{4\,B} + \frac{F\,p''}
   {2\,B\,\left( p + \rho \right) }\nonumber\\
0 & = & 
-\frac{F}{2\,r^2} - \frac{F}{2\,r^2\,B} + \frac{p}{4} + 
  \frac{\rho}{4} - \frac{F'}{2\,r\,B} - 
  \frac{B'\,F'}{8\,{B}^2} + 
  \frac{F\,B'\,p'}{4\,{B}^2\,\left( p + \rho \right) } - 
  \frac{F'\,p'}{4\,B\,\left( p + \rho \right) } + 
  \frac{F\,{p'}^2}{B\,{\left( p + \rho \right) }^2} + 
  \frac{F\,p'\,\rho'}
   {2\,B\,{\left( p + \rho \right) }^2}\nonumber\\
& & + \frac{F''}{4\,B} - 
  \frac{F\,p''}{2\,B\,\left( p + \rho \right)}.\nonumber 
\eea
This set of equations is not linearly independent and one can solve 
$B(r)$ and $B'(r)$Êalgebraically:
\bea{bsols}
B'(r) & = &  
-2\,( 2\,{F}^2\,p' + 
          r^2\,{( p + \rho ) }^2\,
           ( F' - r\,F'' )+ r\,F\,( {p}^2 + {\rho}^2 + 
             p'\,( F' - 
                r^2\,( 2\,p' + \rho' )  )  + 
             \rho\,( -( r\,p' )  + F'' + 
                r^2\,p'' )\nonumber\\
& & +
             p\,( 2\,\rho - r\,p' + F'' + 
                r^2\,p'' )  )  ) \,
        ( r^2\,( p + \rho ) \,{F'}^2\,
           ( p + \rho + r\,p' )  + 
          2\,{F}^2\,( {p}^2 + {\rho}^2 - 
             r^2\,p'\,( 3\,p' + \rho' )  + 
             r^2\,\rho\,p''\nonumber\\
& &  +   p\,( 2\,\rho + r^2\,p'' )  )  + 
          r\,F\,( -( r^2\,F'\,p'\,
                ( 3\,p' + \rho' )  )  + 
             {p}^2\,( 3\,F' - r\,F'' )\nonumber\\
& &   + 
             {\rho}^2\,( 3\,F' - r\,F'' )  + 
             r\,\rho\,( -( r\,p'\,F'' )  + 
                F'\,( 2\,p' + r\,p'' )  )  + 
             p\,( \rho\,
                 ( 6\,F' - 2\,r\,F'' )  + 
                r\,( -( r\,p'\,F'' )  + 
                   F'\,( 2\,p' + r\,p'' )))))\nonumber\\
& &\times ({( p + \rho ) }^3\,
        {( -2\,{F}^2 + 
            r^3\,( p + \rho ) \,F' + 
            r\,F\,( r\,p + r\,\rho - F' - 
               r^2\,p' )  ) }^2)^{-1}\nonumber\\
B(r) & = &  
r\,( -( \frac{r\,{F'}^2\,
               ( p + \rho + r\,p' ) }{p + \rho}
 )  + {F}^2\,( \frac{-2}{r} + 
             \frac{6\,r\,{p'}^2}
              {{( p + \rho ) }^2} + 
             \frac{2\,r\,p'\,\rho'}
              {{( p + \rho ) }^2} - 
             \frac{2\,r\,p''}{p + \rho} )  + 
          F\,( \frac{r\,
                ( p + \rho + r\,p' ) \,F''}{p(
                 r) + \rho}\nonumber\\
& &  + 
             F'\,( -3 + 
                \frac{3\,r^2\,{p'}^2}
                 {{( p + \rho ) }^2} + 
                \frac{r\,p'\,
                   ( -2\,p - 2\,\rho + r\,\rho' ) }\
{{( p + \rho ) }^2} - \frac{r^2\,p''}{p + \rho} \
)  )  )\nonumber\\
& & \times(2\,{F}^2 - 
        r^3\,( p + \rho ) \,F' + 
        r\,F\,( -( r\,p )  - r\,\rho + F' + 
           r^2\,p' ))^{-1}
\eea

\end{widetext}
\end{appendix}

%%%%%%%%%%%%%%%%%%%%%%%%%%%%%%%%%

%%%%%%%%%%%%%%%%%%%%%%%%%%%%%%%%%

\end{document}